\documentclass[traditabstract]{aachanged}
\usepackage[english]{babel}
\usepackage{amsmath}
\usepackage{amssymb}
\usepackage{euscript}
\usepackage{cyr}
\usepackage{epsfig}
\renewcommand{\arraystretch}{1,3}

\begin{document}

\title{HD/H$_2$ Molecular Clouds in the Early Universe:
The Problem of Primordial Deuterium}
\titlerunning{HD/H$_2$ molecular clouds at high redshifts}

\author{S.A. Balashev$^{1}$\thanks{E-mail: balashev@astro.ioffe.ru}, A.V. Ivanchik$^{1,2}$\thanks{E-mail: iav@astro.ioffe.ru} and D.A. Varshalovich$^{1,2}$\thanks{E-mail: varsh@astro.ioffe.ru}}
\authorrunning{Balashev, Ivanchik \& Varshalovich}
\date{Received 15 April, 2010}

\institute{
\it{$^{1}$Ioffe Physical–Technical Institute, ul. Politekhnicheskaya 26, St. Petersburg, 194021 Russia} \\
\it{$^{2}$St. Petersburg State Polytechnical University, ul. Politekhnicheskaya 29, St. Petersburg,
Russia}}

\abstract{ We have detected new HD absorption systems at high redshifts,
z$_{\rm abs}$~=~2.626 and z$_{\rm abs}$~=~1.777, identified in the
spectra of the quasars J\,0812+3208 and Q\,1331+170, respectively. Each
of these systems consists of two subsystems. The HD column densities have
been determined: $\log{\rm N^A_{HD}} = 15.70 \pm 0.07$ for z$_{\rm
A}$~=~2.626443(2) and $\log{\rm N^B_{HD}} = 12.98 \pm 0.22$ for z$_{\rm
B}$~=~2.626276(2) in the spectrum of J\,0812+3208 and $\log{\rm N^C_{HD}}
= 14.83 \pm 0.15$ for z$_{\rm C}$=1.77637(2) and $\log{\rm N^D_{HD}} =
14.61 \pm 0.20$ for z$_{\rm D}$=1.77670(3) in the spectrum of
Q\,1331+170. The measured HD/H$_2$ ratio for three of these subsystems
has been found to be considerably higher than its values typical of
clouds in our Galaxy. We discuss the problem of determining the
primordial deuterium abundance, which is most sensitive to the baryon
density of the Universe $\Omega_b$. Using a well-known model for the
chemistry of a molecular cloud, we have estimated the isotopic ratio
D/H=HD/2H$_2$=$(2.97 \pm 0.55) \times 10^{-5}$ and the corresponding
baryon density $\Omega_bh^2~=~0.0205^{+0.0025}_{-0.0020}$. This value is
in good agreement with $\Omega_bh^2~=~0.0226^{+0.0006}_{-0.0006}$
obtained by analyzing the cosmic microwave background radiation
anisotropy. However, in high-redshift clouds, under conditions of low
metallicity and low dust content, hydrogen may be incompletely
molecularized even in the case of self-shielding. In this situation, the
HD/2H$_2$ ratio may not correspond to the actual D/H isotopic ratio. We
have estimated the cloud molecularization dynamics and the influence of
cosmological evolutionary effects on it. }


\maketitle


\section{Introduction}
\label{introduction} \noindent Being the main coolant of the primordial
gas, molecular hydrogen (both H$_2$ and HD) plays a central role in the
creation of gas condensations and the formation of the first stars in the
post-recombination Universe (see, e.g., Puy et al. 1993; Palla and Galli
1995; Lepp et al. 2002; McGreer and Bryan 2008; Bromm et al. 2009).
Observations of these molecules in high-redshift absorption clouds allow
the physical conditions existed in the early Universe to be determined.

The fact that measuring the HD column density makes it possible to
independently determine the primordial D/H isotopic ratio is important
too (Noterdaeme et al. 2008a; Ivanchik et al. 2010). Among the light
nuclides produced by primordial nucleosynthesis $^2$D, $^3$T, $^3$He,
$^4$He, $^6$Li, $^7$Li, and $^7$Be, the primordial deuterium abundance is
the most sensitive indicator of one of the key cosmological parameters --
the baryon-to-photon ratio, $\eta \equiv n_b/b_{\gamma}$, or the
corresponding baryon density of the Universe $\Omega_b$ (see, e.g.,
Sarkar 1996; Olive et al. 2000; Coc et al. 2005; Fields and Olive 2006).

Until recently, the D/H ratio has been determined mainly only from
H\,{\sc i} and D\,{\sc i} atomic lines in the absorption spectra of
quasars. However, such measurements run into a number of difficulties.
The optical D\,{\sc i} and H\,{\sc i} spectra are almost identical; only
the wavelengths of their lines are shifted by 0.027\%. At the same time,
the number densities of these atoms differ by four or five orders of
magnitude. Therefore, if the H\,{\sc i} column density is low, then the
D\,{\sc i} lines are not seen at all. If, alternatively, the hydrogen
column density is too high, then the H\,{\sc i} lines are saturated,
broadened, and overlap the D\,{\sc i} lines (blend them). Moreover, the
lines identified as D\,{\sc i} ones can in principle be produced by a
small H\,{\sc i} cloud that moves relative to the cloud being studied
with a velocity of $\sim$80 km/s, especially since there are actually
many such clouds moving with different velocities (the socalled
Lyman-$\alpha$ forest) on the line of sight. These factors may also be
responsible for the significant spread in the D/H values obtained by this
method.

No difficulty with the identification of lines arises if the relative
abundance of not the D\,{\sc i} and H\,{\sc i} atoms but the HD and H$_2$
molecules is measured, because their spectra differ significantly and
most narrow absorption lines do not overlap. However, the detection of
molecules at high redshifts is a fairly rare event. For example, H$_2$
molecules are observed only in $\sim$10\% of the DLA systems (Noterdaeme
et al. 2008b) (damped Lyman-$\alpha$ systems -- those with a high atomic
hydrogen column density, N${\rm _{H\,{I}} \gtrsim 10^{20} cm^{-2}}$,
Wolfe et al. 2005) and only 19 high-redshift H$_2$ systems have been
detected to date. Only in three of them have HD lines been detected as
well (the first system is Q\,1232+0815 (Varshalovich et al. 2001;
Ivanchik et al. 2010); the second system is J\,1439+1117 (Noterdaeme et
al. 2008a); and the third system is J\,2123-0050 (Malec et al. 2010)).
The H$_2$ systems are difficult to detect for two reasons. First, the
molecular clouds are compact objects and, hence, the probability of their
falling on the quasar--observer line of sight is low. Second, the H$_2$
absorption lines are very narrow and to detect them high-resolution
spectra with high signal-to-noise ratio are needed. The chance of
obtaining spectra with such a quality increased when large optical
telescopes were put into operation. The HD molecular systems are even
more difficult to observe, because the number density of HD is several
orders of magnitude lower than that of H$_2$. For the HD lines to be seen
in a molecular cloud, the H$_2$ column density must be high enough, $\log
{\rm N_{H_2}} > 19$. This inequality may be considered as a search
criterion for HD systems. Therefore, out of the 16 remaining H$_2$
absorption systems, we gave attention to two systems with $\log {\rm
N_{H_2}} > 19$. These are the absorption systems at z$_{\rm abs}$ = 2.626
in the spectrum of the quasar J\,0812+3208 ($\log {\rm N_{H_2}} = 19.88$,
Jorgenson et al. 2009) and z$_{\rm abs}$ = 1.777 in the spectrum of the
quasar Q\,1331+170 ($\log {\rm N_{H_2}} = 19.65$, Cu iet al. 2005).
Having analyzed these systems, we detected two new HD systems (each of
them consists of two subsystems).

\section{Observational Data}

\begin{figure*}
\centering
\includegraphics[width=167mm,clip]{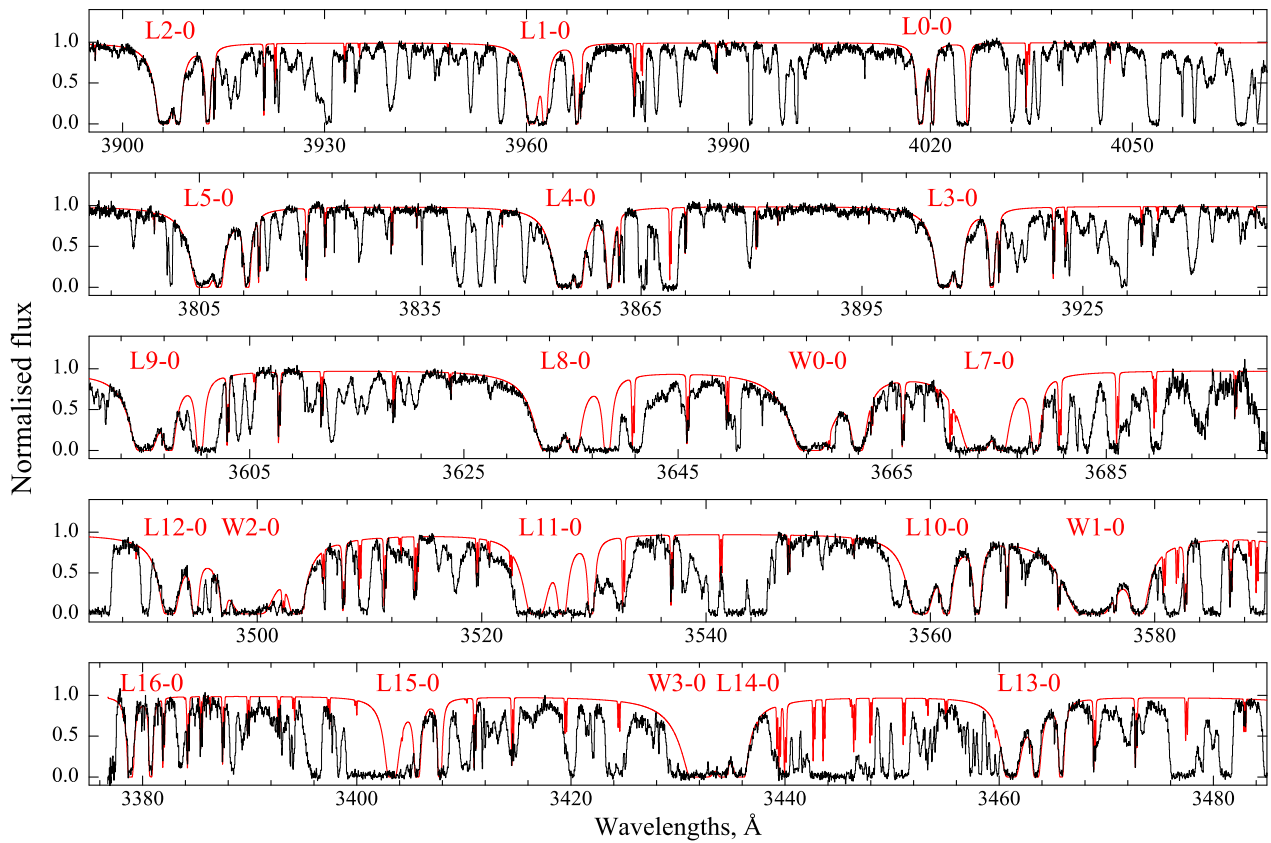}
 \caption{Synthetic H$_2$ spectrum of the absorption system at z = 2.626 fitted
          into the observed spectrum of J\,0812+3208 (HIRES/Keck).}
 \label{H2_fit}
\end{figure*}

\label{observation} \noindent \textbf{J\,0812+3208}. A spectrum of
J\,0812+3208 was taken with the HIRES echelle spectrograph at the Keck I
telescope. This quasar was observed in the wavelength range 3300--6200
\AA\,\,in 2005 under Ellison's and Wolfe's programs (four 1-h and two
1.5-h exposures) and in 2007 and 2008 under Prochaska's program (four 1-h
and four 40-min exposures). The data are in free access from the
Keck/HIRES arhcive\footnote{\tiny
http://www2.keck.hawaii.edu/koa/koa.html}. We reduced and added these
exposures using the MAKEE software package\footnote{\tiny
http://spider.ipac.caltech.edu/staff/tab/makee/index.html} specially
developed by Barlow for Keck/HIRES. The total exposure time was 13.7 h,
which allowed a signal-to-noise ratio of $\sim$45 to be achieved at a
resolution of $\sim$46 000 for the part of the spectrum containing H$_2$
and HD absorption lines.

\textbf{Q\,1331+170}. A high-resolution spectrum of Q\,1331+170 in the
wavelength range 2280--3370 \AA, within which the molecular hydrogen
lines at z~=~1.777 fall, was taken with the STIS spectrograph on the
Hubble Space Telescope (HST) (program 7271 in 1999 and program 9172 in
2002--2003, the principal investigator is Bechtold). The data are in free
access from the HST archive\footnote{\tiny
http://archive.eso.org/archive/hst/}. The exposures were reduced and
added using the CALSTIS procedures of the IRAF software package. The
total exposure time was 14.5 h at a resolution of $\sim$25 000 and the
best signal-to-noise ratio of $\sim$7.

\section{Data Analysis}
\label{Data}
\noindent

\textbf{J\,0812+3208}. A DLA system was identified in the spectrum of
J\,0812+3208 (z$_{\rm em}$~=~2.701) in~2000 (White et al. 2000). The
heavy-element abundance in this system was analyzed by Prochaska et al.
(2003). Analysis of the C\,{\sc I} absorption lines in this system
(Jorgenson et al. 2009) revealed a cold cloud at redshift z$_{\rm
abs}$~=~2.626. The upper limit on the cloud temperature T$< 78$\,K was
estimated from the Doppler parameter of neutral carbon lines. Molecular
hydrogen lines associated with the absorption system were also detected
in the spectrum of this quasar.

\textbf{Q\,1331+170}. A DLA system with $\log {\rm N_{H_2}}$=21.18 was
identified in the spectrum of Q\,1331+170 (Carswell et al. 1975). The
redshift of this system, z~=~1.777, does not allow the part of the quasar
spectrum with molecular hydrogen lines to be obtained by ground based
telescopes. Only HST observations allowed molecular hydrogen lines to be
identified in this system (Cui et al. 2005).

\subsection{H$_2$ Column Densities}

\textbf{J\,0812+3208}. We performed an independent analysis of the
molecular hydrogen absorption system in the spectrum of J\,0812+3208. A
synthetic spectrum was constructed to determine the H$_2$ column density.
The molecular hydrogen lines show the presence of two subsystems in the
spectrum, z$_{\rm A}$~=~2.626443(2) (subsystem A) and z$_{\rm
B}$~=~2.626276(2) (subsystem B), with a relative shift of $\sim$14 km/s.
This structure is most pronounced in the H$_2$ lines associated with the
excited J = 2, 3, 4 rotational levels of the X$^1\Sigma^+_g$, $v = 0$
ground state. The lines associated with the ground states of para- and
orthohydrogen (J = 0 and 1) are strongly saturated; it is hard to resolve
the two subsystems in them and to determine the Doppler parameter b.
Therefore, when determining the H$_2$ column density, we proceeded as
follows. First, we analyzed the J = 2, 3, 4 levels. A local continuum was
constructed for each of the selected lines (a total of $\sim$50 lines
belonging to the Lyman (from L0-0 to L17-0) and Werner (from W0-0 to
W4-0) bands were used). The Doppler parameter was assumed to be the same
for all levels. It should be noted that the parameter b may grow with
increasing rotational level number (see, e.g., Balashev et al. (2009) and
references therein). Here, we disregarded this possibility, because our
main objective was to determine the total hydrogen column density, which
for this system is determined mainly by the ground rotational levels
whose lines are strongly saturated and, as a result, are insensitive to
the parameter b. Once the values of the redshifts and Doppler parameters
for the two subsystems had been obtained, we fixed these values and
constructed a synthetic spectrum for the ground H$_2$ levels. The
synthetic H$_2$ spectrum is shown in Fig.~\ref{H2_fit} and the parameters
obtained from our analysis are presented in Table~\ref{results}. For
subsystem A, our H$_2$ results are in good agreement with those of
Jorgenson et al. (2009). However, the latter authors determined the
Doppler parameter b$_{\rm H_2}$ from the Doppler parameter of neutral
carbon b$_{\rm C\,I}$, while we used it as an independent parameter being
fitted.

For subsystem B, the derived column densities of H$_2$ molecules in the J
= 0 and 1 ground states have large errors, because the column density in
subsystem B is an order of magnitude lower than that in subsystem A.
Therefore, the saturated lines are centered near z$_{\rm A}$ due to the
small relative shift between subsystems A and B, while subsystem B makes
only a slight contribution to the asymmetry of the Lorentz wings.

\begin{figure}
    \includegraphics[width=82mm,clip]{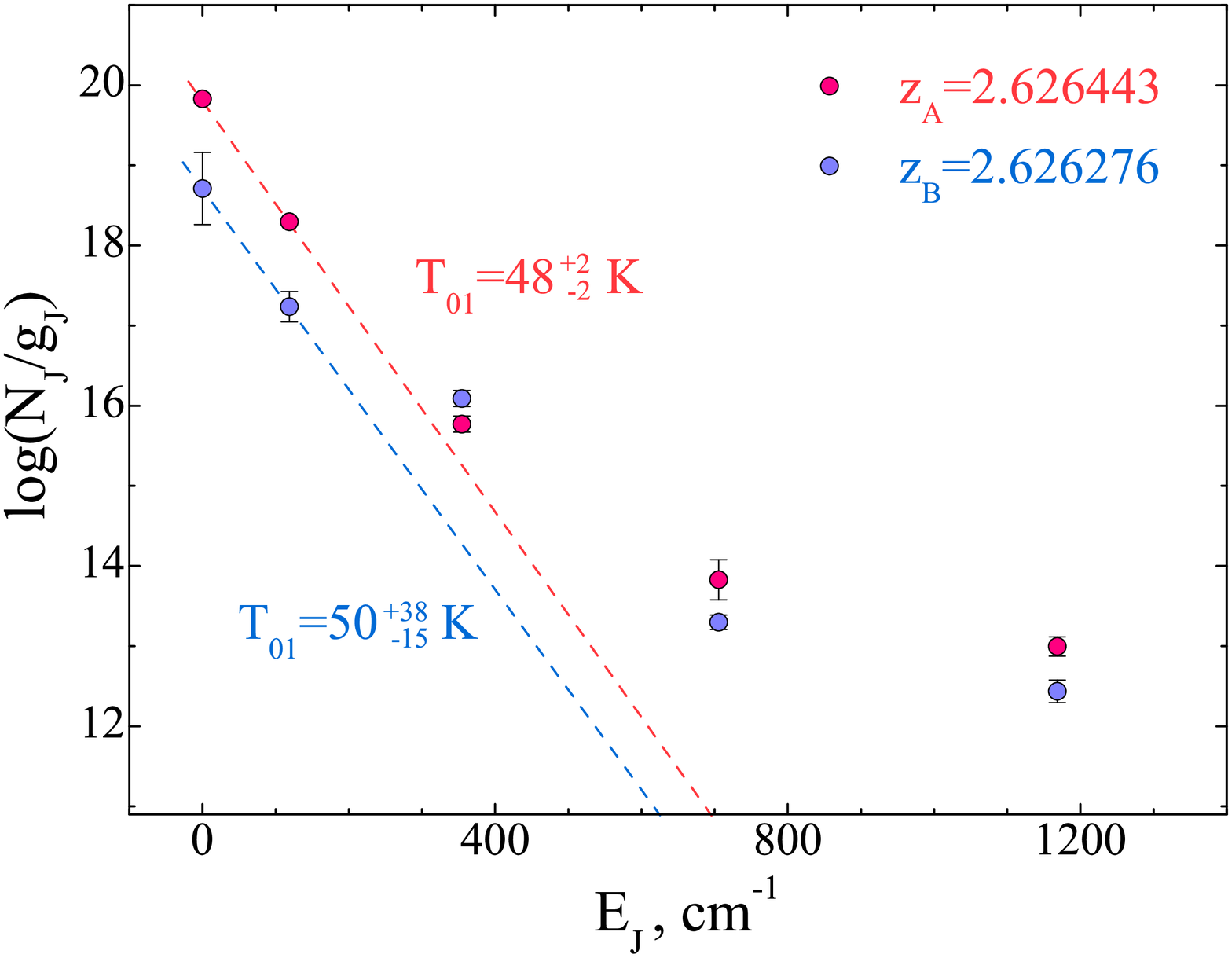}
    \caption{Populations N$_J$ of H$_2$ levels (normalized to their statistical weights g$_J$) versus level energy E$_J$.
             The red and blue circles correspond to subsystems A and B in the spectrum of J\,0812+3208.
             The straight lines corresponds to T$_{01}$, the excitation temperatures of the J = 0 and J = 1 levels.}
                \label{Excitation}
\end{figure}

Fig.~\ref{Excitation} shows the relative populations of H$_2$ rotational
states in the two subsystems. Using the ratio of the ortho-H$_2$ and
para-H$_2$ (J = 0 and 1) column densities, we estimated the gas
temperature in both subsystems: ${\rm T^A_{01}} = 48 \pm 2$\,K and ${\rm
T^B_{01}} = 50^{+38}_{-15}\,$K.

\begin{table*}
 \caption{Results of our analysis of the absorption system at z = 2.626 in the spectrum of J\,0812 + 3208}
   \centering
        \begin{tabular}{c|c|c|c|c|c|c|c}
        \hline \hline
                \multicolumn{4}{c|}{ Subsystem A } & \multicolumn{4}{|c}{ Subsystem B } \\
                \multicolumn{4}{c|}{ $z_{\rm A}=2.626443(2)$ } & \multicolumn{4}{|c}{ $z_{\rm B}=2.626276(2)$ } \\
        \hline
            & & $\log({\rm N})$ & b, km/s & & & $\log({\rm N})$ & b, km/s \\
H$_2$ & ~~J=0~~ & $ ~~19.83 \pm 0.05~~ $ &  $ ~~~0.81 \pm   0.10~~~ $ & H$_2$ & ~~J=0~~ & $ ~~18.71 \pm 0.45~~ $ &    $1.55 \pm   0.11$ \\
            & J=1   & $19.25 \pm 0.02$ &    $0.81 \pm   0.10$ &           & J=1 & $18.19 \pm 0.19$ &    $1.55 \pm   0.11$ \\
            &   J=2 & $16.47 \pm 0.10$ &    $0.81 \pm   0.10$ &           & J=2 & $16.79 \pm 0.10$ &    $1.55 \pm   0.11$ \\
            &   J=3 & $15.15 \pm 0.25$ &    $0.81 \pm   0.10$ &           & J=3 & $14.62 \pm 0.09$ &    $1.55 \pm   0.11$ \\
            &   J=4 & $13.95 \pm 0.12$ &    $0.81 \pm   0.10$ &           & J=4 & $13.39 \pm 0.14$ &    $1.55 \pm   0.11$ \\
            & $\sum_{\rm J}$ & $19.93 \pm 0.04$ &                               &        & $\sum_{\rm J}$ & $18.82 \pm 0.37$ & \\
    & & & & & & & \\
HD      & J=0   & $15.70 \pm 0.07$ &    $0.7 \pm 0.04 $ & HD     & J=0  & $12.98 \pm 0.22$ &    $<3.6^1$ \\
        & J=1   & $13.77 \pm 0.15$ &  $0.7 \pm 0.04 $ & & & &  \\
        & $\sum_{\rm J}$ & $15.71 \pm 0.07$ & & & & & \\
    \hline
  \multicolumn{4}{c|}{} & \multicolumn{4}{|c}{} \\
  \multicolumn{4}{c|}{ ${\rm HD/2H_2 }= 2.97^{+0.52}_{-0.50}\times 10^{-5}$ } & \multicolumn{4}{|c}{ ${\rm HD/2H_2 }= 7.08^{+10.05}_{-4.26}\times 10^{-7}$ } \\
  \multicolumn{4}{c|}{} & \multicolumn{4}{|c}{} \\
    \hline
    \multicolumn{8}{p{15cm}}{$^1$ - The observed lines are insensitive to the parameter b,
                                   because they are located at the linear part of the curve of growth.} \\

    \label{results}
  \end{tabular}
\end{table*}

\textbf{Q\,1331+170}. In contrast to the analysis of the H$_2$ absorption
system in Cui et al. (2005), where the authors fitted only one component
(z~=~1.776553(3)) into the spectrum, we constructed a synthetic spectrum
by assuming the absorption system to consist of two components (z$_{\rm
C}$~=~1.77637(2) and z$_{\rm D}$~=~1.77670(3)). Two components with a
relative shift of $\sim$35 km/s are clearly seen in the H$_2$ lines
originating from the J = 4 rotational level (Fig.~\ref{H2_fit_2}) and in
the HD lines we identified (see Fig.~\ref{HD_fit_2} below). The redshifts
of both subsystems were determined by analyzing the H$_2$ (J~=~4) lines.
Since the lines associated with the J = 0 and 1 levels are strongly
saturated for both subsystems, the column density of molecules at these
levels can be obtained with a good accuracy. This cannot be said for the
lines associated with the J~=~2 and 3 levels, which fall into the
logarithmic part of the curve of growth. As a result, the total molecular
hydrogen column density for the two subsystems is $\log {\rm N^C_{H_2}} =
19.43 \pm 0.10$ and $\log {\rm N^D_{H_2}} = 19.39 \pm 0.11$.

\begin{figure*}
\hspace{7mm}
\includegraphics[width=147mm,clip]{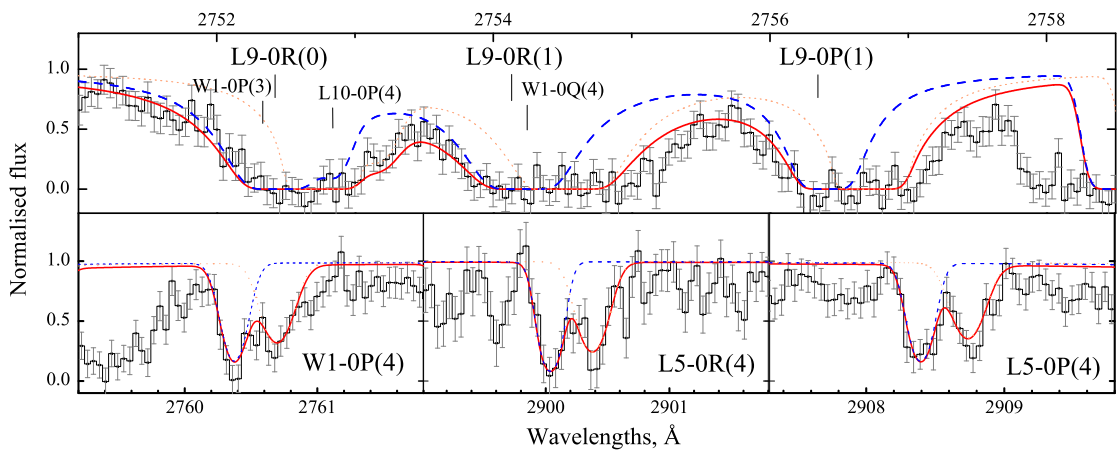}
 \caption{Synthetic H$_2$ spectrum of the absorption system at z = 1.777 fitted into the observed spectrum of Q\,1331+170
(STIS/HST). The two components of the synthetic spectrum are indicated by the dotted and dashed lines. These components
are well resolvable for the lines originating from the J = 4 rotational level (bottom).}
 \label{H2_fit_2}
\end{figure*}

\subsection{HD Column Densities}

\begin{figure*}
\centering
\includegraphics[width=144mm,clip]{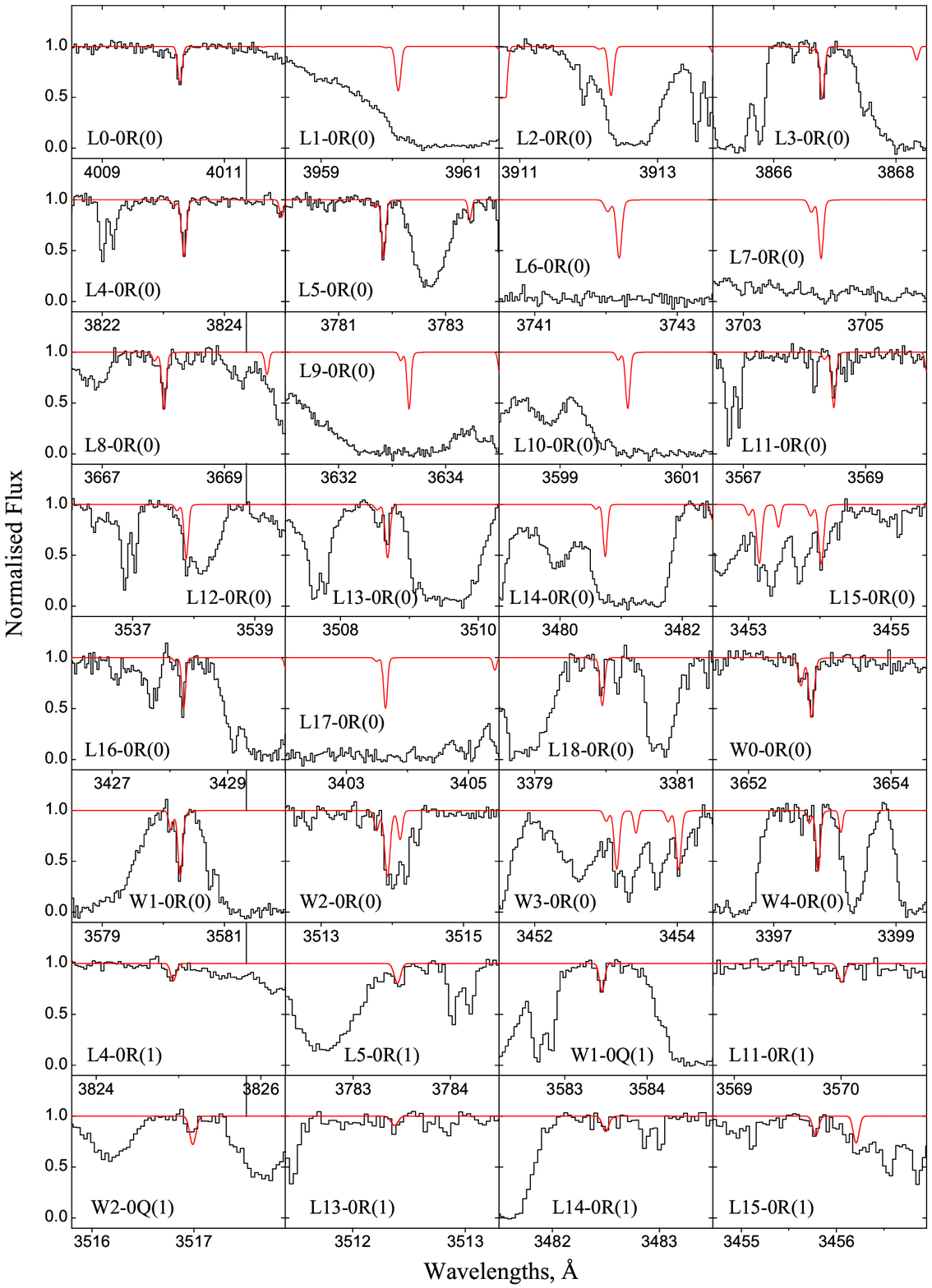}
 \caption{Synthetic HD molecular spectrum of the absorption system at z~=~2.626 in the observed spectrum of J\,0812+3208.
Two components are seen for the transitions from the J = 0 ground state.
The R(1), P(1), and Q(1) lines of the transitions from the excited J = 1
rotational level are also shown.}
 \label{HD_fit}
\end{figure*}

\textbf{J\,0812+3208}. To determine the HD column density, the Doppler
parameter, and the redshift, we constructed a synthetic spectrum for each
of the two subsystems (A and B). The parts of the quasar spectrum
normalized to the local continuum and the synthetic spectrum for HD lines
are shown in Fig.~\ref{HD_fit}. In subsystem A, we identified the HD
lines corresponding to the transitions not only from the J = 0 ground
state but also from the J = 1 rotational level. This allowed us to
estimate the number density in the cloud, n $= 54^{+36}_{-22}\,{\rm
cm}^{-3}$ (for more detail, see below). As far as we know, this is the
first reliable detection of HD lines associated with the J = 1 level.
This proved to be possible due to the high HD column density and good
spectrum quality.

The HD molecular lines of the Lyman and Werner series are present in the
spectrum up to L18-0 and W4-0 inclusively. In general, the HD wavelengths
and oscillator strengths for L$\nu''-0$ and W$\nu''-0$ lines with large
$\nu''$ are not known well enough. The laboratory wavelengths were
measured with a high accuracy, $\delta\lambda/\lambda \sim 5 \times
10^{-8}$, only for the bands up to L9-0 and W0-0 (Ivanov et al. 2008).
For shorter wavelength transitions, we used the laboratory data (Ivanov
et al. 2010) kindly provided to us by Prof. W. Ubachs. The oscillator
strengths were taken from the calculations by Abgrall and Roueff (2006);
the wavelengths are also given there. However, comparison of the
theoretical (Abgrall and Roueff 2006) and laboratory (Ivanov et al. 2008)
values showed that the calculated wavelengths for some lines differ
significantly from the laboratory ones. For example, the calculated
wavelength of the L11-0R(0) line differs from the laboratory value by
$\sim$8 km/s and does not fit into the general analysis. Using the data
from Ivanov et al. (2010), we were able to determine the HD column
density with a good accuracy.

\begin{figure}
\centering
    \includegraphics[width=74mm,clip]{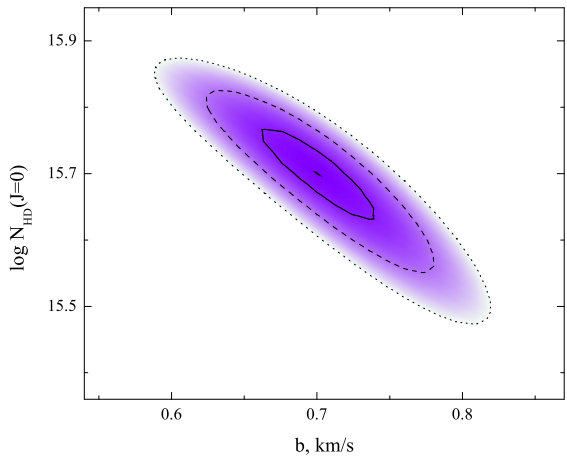}
        \caption{Confidence regions of the parameters N$_{\rm HD}$ and b$_{\rm HD}$ for the J = 0 level of subsystem A in the spectrum of J\,0812+3208. The minimum corresponds to the column density $\log {\rm N_{HD}} = 15.70$ and the Doppler parameter b$_{\rm HD}$ = 0.7 km/s.}
        \label{HD_J0}
\end{figure}

The results of our determination of the parameters for the identified HD
molecular lines in the spectrum of J\,0812+3208 are presented in
Table~\ref{results}. The redshifts of the HD lines in subsystems A and B
coincide, within the error limits, with those determined by analyzing
H$_2$ lines. To determine the HD (J = 0) column density in subsystem A,
we chose the R(0) lines of the following bands: L0-0, L3-0, L4-0, L5-0,
L8-0, L11-0, L18-0, W0-0, W4-0. The 1, 2, and 3$\sigma$ confidence
regions for N$_{\rm HD}$ and b$_{\rm HD}$ are shown in Fig.~\ref{HD_J0}.
We determined the HD column density in subsystem A, $\log {\rm
N_{HD}}$~=~$15.71 \pm 0.07$. This is the highest value of N$_{\rm HD}$
measured in the quasar spectra. The derived Doppler parameter b$_{\rm HD}
= 0.70 \pm 0.04\,\,{\rm km/s}$ agrees (within the error limits) with
b$_{\rm C\,I}  = 0.33 \pm 0.05\,\,{\rm km/s}$ and b$_{\rm H_2} = 0.81 \pm
0.10\,\,{\rm km/s}$ (if the thermal broadening is dominant, then the
Doppler parameters are related as the square root of the ratio of the
element masses).

\textbf{Number density in the subsystem J\,0812+3208A}. The relative
populations of the HD J = 0 and J = 1 rotational levels determined in the
absorption subsystem J\,0812+3208A allow the number density of the
molecular cloud to be estimated. This requires considering the level
population balance equations. The J = 1 level can be populated by the
direct radiative transition from the J = 0 level, radiative pumping
through excited electronic states, and collisions. Typically, the
particle number density in a diffuse cloud is n$\sim 10-500\,\,{\rm
cm}^{-3}$. At such number densities and the Galactic mean intensity of
ultraviolet radiation, collisions will be the dominant J = 1 level
population channel for the bulk of the cloud. Indeed, at such a high
measured column density in the cloud, $\log {\rm N_{HD}} = 15.7$, the
rate of the 0$\to$1 population by radiative pumping inside the cloud
falls by more than two orders of magnitude through selfshielding. The
direct radiative 0$\to$1 population can also be neglected. The dominant
channel of the 1$\to$0 transition is spontaneous relaxation characterized
by the radiative transition probability A$_{10}$. Thus, we can estimate
the number density from the formula
$$
    n = \frac{N_{J=1}}{N_{J=0}}\frac{A_{01}}{C_{01}}
$$

The main collisional parter of HD for a molecular cloud will be H$_2$;
the values of the collisional probability coefficient C$_{01}$ were taken
from Flower et al. (2000). Using a temperature estimate in the cloud of
T$^A_{01} = 48 \pm 2\,\,{\rm K}$, we obtained the following estimate for
the number density in the cloud: n$ = 54^{+36}_{-22}\,\,{\rm cm}^{-3}$ it
agrees with typical values for diffuse clouds (Rachford et al. 2002).
Note, however, that if the cloud is not molecularized, then the main
collisional partner is atomic hydrogen and the number density estimate
doubles.

\begin{figure*}
\centering
\includegraphics[width=170mm,clip]{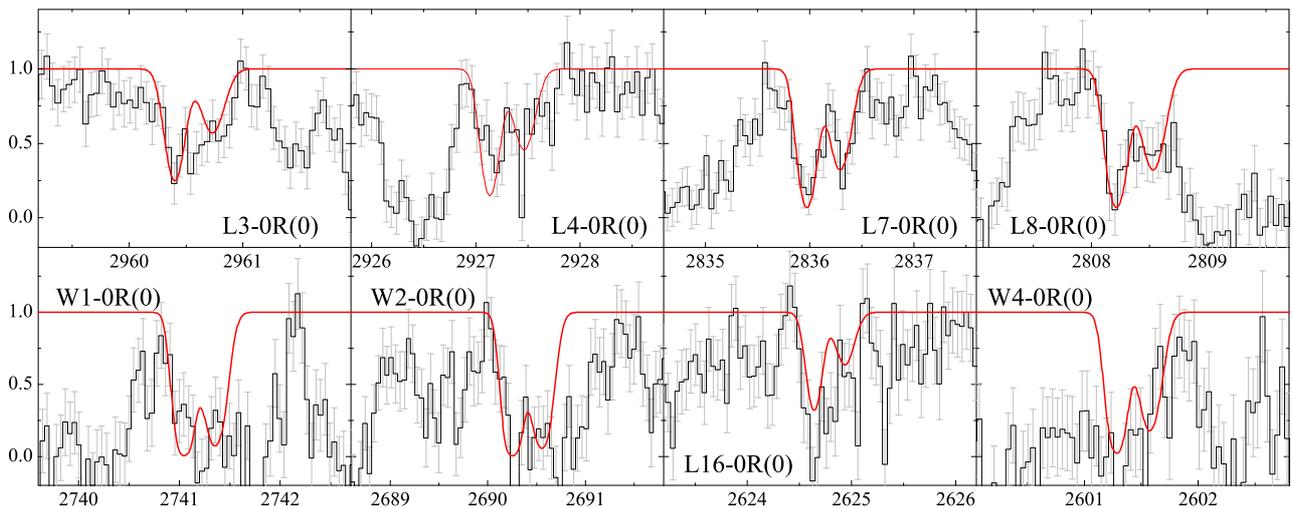}
 \caption{Synthetic HD spectrum of the absorption system at z = 1.777 fitted into the observed spectrum of Q\,1331+170. The R(0) lines of the transitions from the J = 0 ground state, in which two components are clearly seen, are shown.}
 \label{HD_fit_2}
\end{figure*}

\textbf{Q1331+170}. The HD molecular lines are present in the spectrum up to the L18-0 Lyman and W4-0 Werner bands, respectively (some of these lines are shown in Fig.~\ref{HD_fit_2}). Because of the low signal-to-noise ratio, we were able to use only the L3-0R(0), L4-0R(0), L7-0R(0), and L8-0R(0) lines to determine the HD parameters. Two subsystems whose redshifts coincided, within the error limits, with those determined by analyzing H$_2$ are seen in the chosen lines.
The HD column densities determined by synthetic spectrum construction were found to be $\log {\rm N^C_{HD}} = 14.83 \pm 0.15$ and $\log {\rm N^D_{HD}} = 14.61 \pm 0.20$.

\subsection{HD/2H$_2$ Ratio}
\begin{figure*}
\centering
\includegraphics[width=157mm,clip]{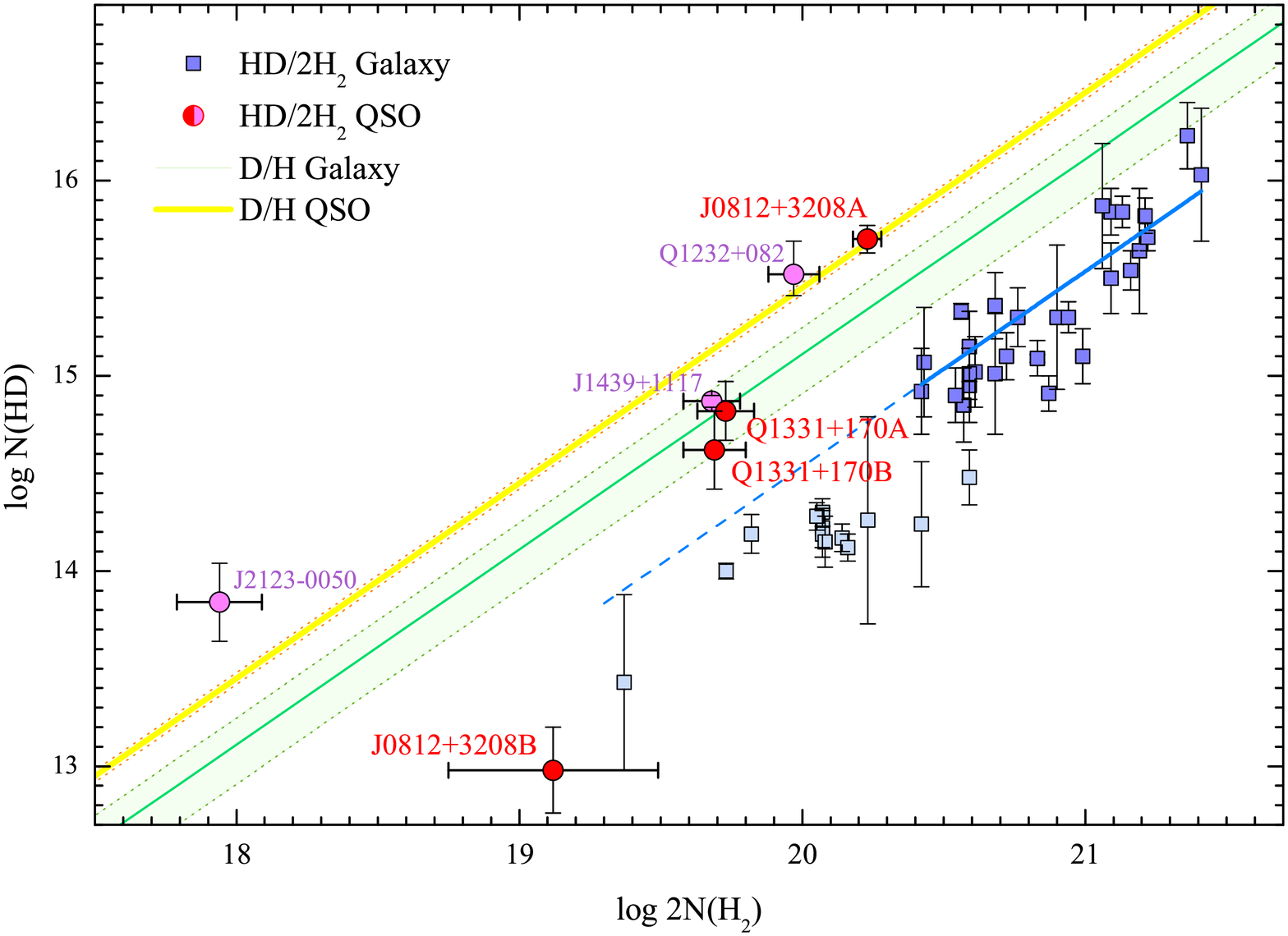}
 \caption{D\,{\sc i}, H\,{\sc i} and HD, H$_2$ column density measurements. The circles indicate the results of HD and H$_2$ measurements in the quasar spectra (the red circles correspond to the data obtained in this paper); the squares represent the results of HD and H$_2$ measurements in our Galaxy (Snow et al. 2008; Lacour et al. 2005); the blue and light blue squares indicate the points for which the HD column density is higher and lower than $\log {\rm N_{HD}}$ = 15, respectively. The solid straight line indicates the mean HD/2H$_2$ ratio determined from points with $\log {\rm N_{HD}} > $ 15; the green stripe indicates the results of H\,{\sc i} and D\,{\sc i} measurements taken from Linsky et al. (2006) (the mean D/H ratio for these values is indicated by the dotted line). The D/H ratio in the quasar spectra measured from H\,{\sc i} and D\,{\sc i} lines (Pettini et al. 2008) is indicated by the yellow stripe.}
 \label{HD_H2}
\end{figure*}

Until now, HD lines (at high redshifts) have been identified only in
three of the 19 H$_2$ absorption systems. Figure~\ref{HD_H2} shows the
HD, H$_2$ and D\,{\sc i}, H\,{\sc i} column densities measured in
interstellar clouds of our Galaxy and absorption systems of quasars.
Determining the column densities of HD and H$_2$ or D\,{\sc i} and
H\,{\sc i} at high redshifts allows the primordial D/H isotopic ratio
and, hence, $\Omega_b$ to be estimated. The oblique straight lines in
Fig.~\ref{HD_H2} indicate the column densities that correspond to the
mean D/H and HD/2H$_2$ ratios for the Galactic and extragalactic systems.
In our Galaxy, the D/H ratio measured from the column densities of atomic
D\,{\sc i} and H\,{\sc i} lines (Linsky et al. 2006) is systematically
lower than its mean value obtained by analyzing the quasar spectra
(Pettini et al. 2008). In principle, this can be explained by the
astration of deuterium in stars. We also see that there is a significant
difference between the values of D/H and HD/2H$_2$ measured in our
Galaxy. This may be because in contrast to H$_2$, the HD molecules are
not always shielded from ultraviolet radiation and, hence, deuterium is
molecularized to a lesser extent. Another explanation for the shortage of
the HD molecular fraction can be complex chemistry of molecular clouds,
where deuterium can effectively enter into other, more complex molecules,
H$_2$O, NH$^3$,  HCN, polyaromatic hydrocarbons, etc. No significant
difference is observed for the D/H ratios in the spectra of quasars
measured from the molecular and atomic components (Fig.~\ref{HD_H2}).
However, the statistics for quasar spectra is much poorer than that for
systems in our Galaxy.

\textbf{J\,0812+3208}. The ratios of the HD and H$_2$ column densities in
the spectrum of J\,0812+3208 were found to be HD/2H$_2 =
2.97^{+0.52}_{-0.50} \times 10^{-5}$ and HD/2H$_2 = 7.08^{+10.05}_{-4.26}
\times 10^{-7}$ for the absorption systems A and B, respectively. The
HD/2H$_2$ ratio in subsystem B is definitely unsuitable for the D/H
estimate, because the HD column density in this subsystem ($\log {\rm
N_{HD}}$~=~12.98) is two orders of magnitude lower than the value
necessary for the self-shielding of molecules and, as a result, deuterium
may be incompletely molecularized. According to the calculations by Le
Petit et al. (2002), if HD and H$_2$ are self-shielded from ultraviolet
radiation, then both components are completely molecularized and the
universal ratio D/H~=~HD/2H$_2$ is established in the cloud. In our case,
only subsystem A can be completely shielded in HD, because it has a
relatively high HD column density: $\log {\rm N_{HD}}$~=~15.7. Besides,
this subsystem has a low metallicity (relative to the solar one),
$\lesssim -1.0$ (Prochaska et al. 2003). This means a low level of
deuterium astration in stars and makes it possible to estimate the
primordial D/H ratio. On the other hand, a low metallicity also means a
low dust content; since molecular hydrogen is formed mainly on dust, this
leads to a lower formation rate of H$_2$ and HD and, in contrast to the
calculations by Le Petit et al. (2002), the cloud molecularization may be
incomplete. Since there is an additional, fairly efficient formation
channel for HD, H$_2$+D$^+$ $\to$ HD + H$^+$, at a low H$_2$ formation
rate on dust this can lead to a significant increase in the HD/2H$_2$
ratio and it will not correspond to the true D/H isotopic ratio (see,
e.g., Ferlet et al. 2000). Besides, at a low dust content, the process of
gas molecularization in the cloud slows down. Therefore, the matter in
the observed clouds may not have time to reach a steady state in chemical
equilibrium and the HD/2H$_2$ ratio can then be different for clouds of
different ages (see the Section~\ref{MolModel} and Fig.~\ref{Model}).

\textbf{Q\,1331+170}. The ratios of the HD and H$_2$ column densities in
the spectrum of Q\,1331+170 were found to be HD/2H$_2 =
1.24^{+0.60}_{-0.40} \times 10^{-5}$ and HD/2H$_2 = 0.83^{+0.54}_{-0.33}
\times 10^{-5}$ for the absorption systems C and D, respectively. Note
that the large errors in the parameters compared to the previous system
are attributable to a poorer spectrum quality.

\subsection{Peculiarities of the HD/H$_2$ Absorption Systems}

\textbf{J\,2123-0050}. According to Malec et al. (2010), the absorption
system in the spectrum of J\,2123-0050 has the highest HD/2H$_2$ ratio,
$\sim 7 \times 10^{-5}$. However, in our opinion, this system requires an
additional study of the column densities and its physical parameters,
because the ortho-to-parahydrogen ratio obtained by these authors, $\sim$
16:1, exceeds considerably the ratio 3:1, appears very strange, and has
never been observed before.

\textbf{J\,1439+1117}. The metallicity of this system is nearly solar and
HD/2H$_2 = 1.5\times 10^{-5}$ is close to the D/H ratio measured in our
Galaxy (Noterdaeme et al. 2008a).

\textbf{Q\,1232+0815, J\,0812+3208}. This system has a low metallicity,
$\sim -1.5$ (Balashev et al. 2010; Prochaska et al. 2003), and high HD
column densities. This makes the derived values for these HD/2H$_2$
systems a good estimate of the primordial D/H ratio.

\textbf{Q1331+170}. The metallicity of this system is also low, $\sim
-1.5$ (Prochaska and Wolfe 1999). However, incomplete shielding in HD is
quite possible for the two derived HD/H$_2$ systems, because the HD
column density is low. Besides, the spectrum has a low signal-to-noise
ratio and, hence, the estimate of the primordial D/H ratio based on these
systems is ambiguous.

The results for all the systems are summarized in Table~\ref{HDtoH2}.

\begin{table*}
    \caption{HD/H$_2$ absorption systems in the quasar spectra. }
  \centering
     \begin{tabular}{c|c|c|c|c|c}
     \hline
     QSO & z$_{abs}$ & N$_{{\rm H}_2}$, cm$^{-2}$ & N$_{\rm ND}$, cm$^{-2}$ & Instrument & References$^1$ \\
     \hline \hline
     & & & & & \\
     Q\,1232+0815  & 2.33714(3) & $4.78^{+0.96}_{-0.96}\times10^{19}$ & $3.39^{+1.6}_{-0.8}\times10^{15}$ & UVES & 1, 2 \\
     & & & & & \\
     J\,1439+1117$^2$  & 2.41837 & $2.40^{+0.49}_{-0.62}\times10^{19}$ & $7.18^{+0.41}_{-0.44}\times10^{14}$ & UVES & 3 \\
     & & & & & \\
     J\,2123+0050 & 2.0593276(5) & $3.68^{+0.33}_{-0.30}\times10^{17}$ & $5.89^{+0.42}_{-0.39}\times10^{13}$ & KECK & 4 \\
     & & & & & \\
     Q\,0812+3208  & 2.626443(2) & $8.54^{+0.83}_{-0.76}\times10^{19}$ & $5.07^{+0.88}_{-0.75}\times10^{15}$ & KECK & 5 \\
     & & & & & \\
     Q\,0812+3208 & 2.626376(2) & $6.74^{+9.36}_{-3.92}\times10^{18}$ & $9.55^{+6.30}_{-3.80}\times10^{12}$ & KECK & 5 \\
     & & & & & \\
     Q\,1331+170  & 1.77637(2) & $2.66^{+0.66}_{-0.53}\times10^{19}$ & $6.61^{+2.73}_{-1.93}\times10^{14}$ & HST & 5 \\
     & & & & & \\
     Q\,1331+170  & 1.77670(2) & $2.45^{+0.69}_{-0.54}\times10^{19}$ & $4.07^{+2.38}_{-1.50}\times10^{14}$ & HST & 5\\
     & & & & & \\
     \hline
     \multicolumn{6}{p{13.7cm}}{$^1$ Refs.: 1 — Varshalovich et al. (2001), 2 — Ivanchik et al. (2010), 3 — Noterdaeme et al. (2008a),
     4 — Malec et al. (2010), 5 — this paper.

     $^2$ In this system, three components are seen in the HD lines; however, since the H$_2$ column densities cannot be determined for these components, the authors provide only an integrated value.} \\
    \label{HDtoH2}
    \end{tabular}
\end{table*}

\section{Cloud Molecularization Model}

\begin{figure*}
\centering
    \includegraphics[width=140mm,clip]{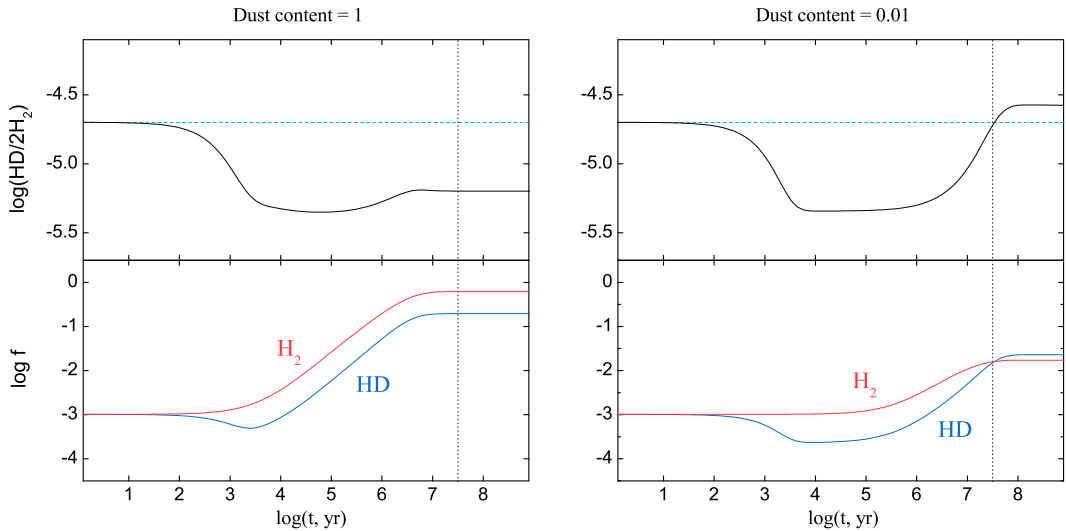}\\
        \caption{ Time variation of the HD/2H$_2$ abundance ratio (top panels) and the molecular fraction (bottom panels) during the evolution of an interstellar cloud with a number density n$ = 100 {\rm cm}^{-3}$ and T = 80 K. The calculations are presented for the Galactic dust content (left panels) and 0.01 of the Galactic dust content (right panels). The vertical line marks the characteristic dynamical lifetime of clouds, $\sim 3 \times 10^{7}$ yr.}
                \label{Model}
\end{figure*}

\label{MolModel} To estimate the influence of incomplete molecularization
on the HD/H$_2$ ratio, we calculated the chemical evolution of an
HD/H$_2$ molecular cloud (this model will be presented in detail in
Balashev et al. (2010)). Figure~\ref{Model} presents the results of our
calculations of the chemical evolution of a typical interstellar cloud
with a temperature T = 80 K and a total hydrogen number density n = 100
cm$^{-3}$. We consider a cloud composed of H, He, D, their ions,
electrons, H$_2$ and HD molecules, and some amount of dust on which these
molecules are efficiently formed. The chemical reaction rates were taken
from Stancil et al. (1998) and Glover and Jappsen (2007). We also took
into account more complex molecules, H$^{+3}$, H$_2$D$^+$, HeH$^+$,
HeD$^+$, etc., but their influence on the HD/H$_2$ evolution of interest
to us proved to be insignificant in the case of a diffuse molecular
cloud. The cloud was assumed to be in the field of external ultraviolet
radiation, which leads to an efficient dissociation of molecules in the
cloud surface layers. Its intensity was taken to be equal to the mean
intensity of ultraviolet radiation in our Galaxy. The function of cloud
self-shielding from ultraviolet radiation was taken from Draine and
Bertoldi (1996). We also assumed that the cloud was irradiated by cosmic
rays, which leads to gas ionization that provides ion–molecular
reactions. The cosmic-ray ionization rate was taken to be equal to its
Galactic mean value. The dust content was taken to be 1 (Fig. 8, left) or
0.01 (Fig. 8 right) of the mean dust content in our Galaxy. These values
were chosen so as to qualitatively explain the difference between the
observed clouds in our Galaxy and high-redshift DLA systems. The total
hydrogen column density in the system N$_{\rm tot}$ = N$_{\rm H\,{I}}$ +
2N$_{\rm H_2}$ was chosen from the condition N$_{\rm H_2} \simeq 10^{20}
{\rm cm}^{-2}$, which corresponds to the column density for the
absorption subsystem A of the quasar J\,0812+3208.

We see that this model can explain the observational data, provided that
the cloud age is greater than 10$^7$ yr. At the Galactic dust content,
the cloud reaches almost complete molecularization in H$_2$, f$_{\rm H_2}
= \frac{2N_{H_2}}{2N_{H_2} + N_{H\,{I}}} \approx 1$ (Figs. 8 bottom
left); the HD molecular fraction (f$_{HD} = \frac{N_{HD}}{N_{HD} +
N_{D\,{I}}}$) is lower, because HD is shielded more weakly against
ultraviolet radiation than H$_2$. The situation is different at a low
dust content. The hydrogen molecular fraction is low (f$_{\rm H_2}$ ,
f$_{HD} < 0.05$); the ion reaction of HD formation becomes more efficient
than the formation of HD on dust and, as a result, the molecular fraction
of HD can be higher than that of H$_2$ (Figs. 8 bottom right).

It is important to note that HD/H$_2$ depends significantly on the cloud
age at a low dust content (Figs. 8 top panels). In addition, the HD/H$_2$
ratio also depends on other factors, namely, the ultraviolet radiation
background, the cosmic-ray intensity, the gas number density, and the
total column density. Therefore, the HD/2H$_2$ ratio may not correspond
to the actual D/H isotopic ratio and its estimation requires developing a
model for the cloud molecularization dynamics (Balashev et al. 2010).
Note that for the two most saturated (in HD) systems (J\,0812+3208A and
Q\,1232+082, see Fig. 7), the HD/2H$_2$ ratios turned out to be
coincident (within the error limits) and close to the primordial D/H
determined by analyzing the cosmic microwave background radiation
anisotropy (Komatsu et al. 2010). It is hoped that increasing the
statistics of high-redshift molecular clouds will clarify this situation.

In molecular cloud models, the D/HD transition is assumed to occur at
larger penetration depths of the emission into the cloud than the H/H2
transition (see, e.g., Le Petit et al. 2002). However, under conditions
of low dust content, a situation where the D/HD transition in the cloud
will begin more early than the H/H$_2$ transition is possible. This will
occur via the ion reaction of HD formation under incomplete H$_2$
molecularization conditions.

At a low dust content and high column densities, N$_{\rm H_2} \gtrsim
10^{21} {\rm cm}^{-2}$, on long time scales ($\gtrsim 10^{8}$ yr), the
cloud reaches complete molecularization in H$_2$ and HD and the HD/2H$_2$
ratio will corresponds to the D/H ratio. However, there will exist a long
time interval ($\sim 10^{7}$ yr) in the cloud evolution in which the
HD/2H$_2$ ratio will exceed considerably the D/H ratio.

\section{Conclusions}

We identified HD molecular lines in the absorption systems at z$_{\rm
abs}$~=~2.626 and z$_{\rm abs}$~=~1.777 in the spectra of the quasars
J\,0812+3208 and Q\,1331+170, respectively. Each system consists of two
clearly resolvable components. The HD and H$_2$ column densities were
determined for all four subsystems (Table 2).

Since only one of the four identified subsystems, at z = 2.626443(2), in
the spectrum of J\,0812+3208 has an HD column density exceeding its
critical value for self-shielding, one would expect the HD/2H$_2$ ratio
in the steady-state model of a completely molecularized cloud to
characterize the primordial D/H isotopic ratio. Under these conditions,
D/H$=(2.97^{+0.52}_{-0.50})\times10^{-5}$, which corresponds to the
baryonic matter density $\Omega_{\rm b}h^2=0.0205^{+0.0025}_{0.0020}$.
This value is in good agreement with $\Omega_{\rm
b}h^2=0.0226^{+0.0006}_{0.0006}$ obtained by analyzing the cosmic
microwave background radiation anisotropy (Komatsuet al. 2010).

However, the cloud molecularization dynamics and the influence of
cosmological evolutionary factors related to a change in dust content,
ionizing radiation background, and cosmic-ray intensity on it can lead to
a systematic shift in the estimate of the primordial D/H isotopic ratio.

We made preliminary estimates of the effects related to the cloud
molecularization dynamics in the early Universe. Their allowance was
shown to be necessary to properly determine the primordial D/H ratio.

Increasing the statistics of HD/H2 molecular systems will allow progress
to be made both in understanding the chemical and isotopic evolution of
molecular clouds as well as for determining the primordial D/H isotopic
ratio. Significant progress in solving this problem may be expected when
future extremely telescopes will be put into operation.

\section*{Acknowledgments}

We thank Prof. W. Ubachs who kindly provided the HD wavelengths. This
work was supported by the Russian Foundation for Basic Research (project
no. 08-02-01246a) and the Program of the Russian President for Support of
Leading Scientific Schools (NSh-3769.2010.2).

\begin{figure*}
\centering
\includegraphics[width=164mm,clip]{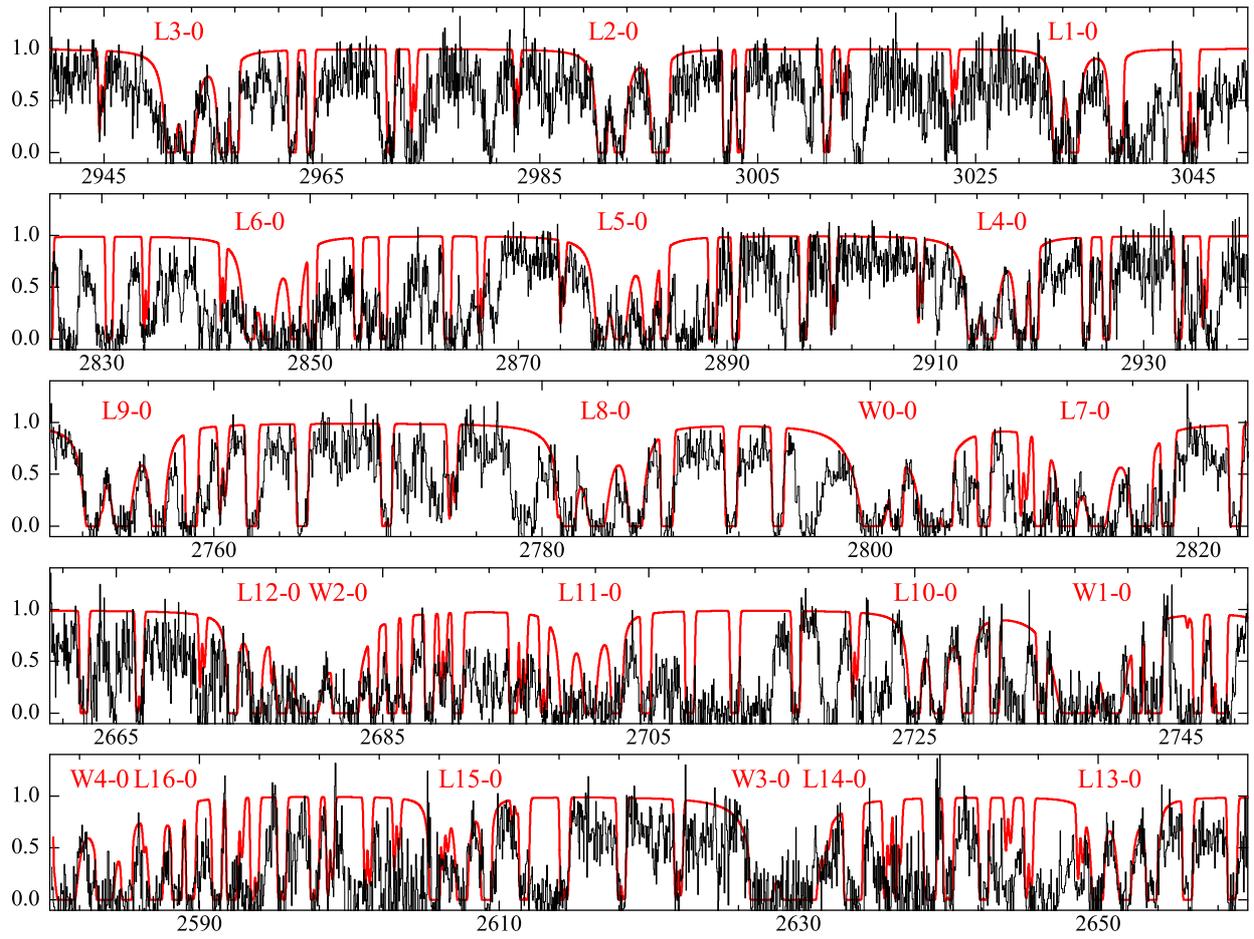}
 \caption{Synthetic H$_2$ spectrum of the absorption system at z = 1.777 fitted
          into the observed spectrum of Q\,1331+170 (STIS/HST).}
 \label{H2_fit_2_all}
\end{figure*}

\end{document}